# Regional impacts poorly constrained by climate sensitivity

Authors: Ranjini Swaminathan[1,*,#], Jacob Schewe[2,*,#], Jeremy Walton[3], Klaus Zimmermann[4], Colin Jones[5], Richard A. Betts[3,6], Chantelle Burton[3], Chris D. Jones[3,7], Matthias Mengel[2], Christopher P.O. Reyer[2], Andrew G. Turner[8], Katja Weigel[9,10]

Affiliations:

[1]National Centre for Earth Observation and Department of Meteorology, University of Reading, Reading, RG6 6ES, United Kingdom

[2]Potsdam Institute for Climate Impact Research, Member of the Leibniz Association, 14473 Potsdam, Germany

[3]Met Office Hadley Centre for Climate Science and Services, Exeter, EX1 3PB, United Kingdom

[4]Swedish Meteorological and Hydrological Institute, Rossby Centre, SE-601 76 Norrköping, Sweden

[5]National Centre for Atmospheric Science and School of Earth and Environment, University of Leeds, Leeds, LS2 9TJ, United Kingdom

[6]Global Systems Institute, University of Exeter, Exeter, EX4 4QE, United Kingdom

[7]School of Geographical Science, University of Bristol, Bristol, BS8 1HB, United Kingdom

[8]National Centre for Atmospheric Science and Department of Meteorology, University of Reading, Reading, RG6 6ES, United Kingdom

[9]University of Bremen, Institute of Environmental Physics (IUP), Bremen, Germany

[10]Deutsches Zentrum für Luft- und Raumfahrt (DLR), Institut für Physik der Atmosphäre, Oberpfaffenhofen, Germany

[#]Co-lead authors

[*]Corresponding authors

# Abstract

Climate risk assessments must account for a wide range of possible futures, so scientists often use simulations made by numerous global climate models to explore potential changes in regional climates and their impacts. Some of the latest-generation models have high effective climate sensitivities (EffCS). It has been argued these "hot" models are unrealistic and should therefore be excluded from analyses of climate change impacts. Whether this would improve regional impact assessments, or make them worse, is unclear. Here we show there is no universal relationship between EffCS and projected changes in a number of important climatic drivers of regional impacts. Analysing heavy rainfall events, meteorological drought, and fire weather in different regions, we find little or no significant correlation with EffCS for most regions and climatic drivers. Even when a correlation is found, internal variability and processes unrelated to EffCS have similar effects on projected changes in the climatic drivers as EffCS. Model selection based solely on EffCS appears to be unjustified and may neglect realistic impacts, leading to an underestimation of climate risks.

# Introduction

The use of climate projections for impact assessment often exploits numerous climate and Earth system models (ESMs). The Coupled Model Intercomparison Project[1] (CMIP) provides underpinning simulations to the Intergovernmental Panel on Climate Change (IPCC) analysis of climate projections[2], changes in climate impact drivers[3], and assessment of impacts and vulnerability[4]. Using a diverse set of ESMs, the CMIP ensemble aims to span the uncertainty range of future projections, including potential high impact, low-likelihood events[5]. It is important that CMIP models represent plausible future evolutions of the climate system and thus perform well against observations[6]. A range of studies have shown improvement over generations of models, including from CMIP5 to CMIP6[7]. We therefore expect CMIP6 to represent an advance for use in impact studies.

One widely discussed change from CMIP5 to CMIP6 is an increased range in model effective climate sensitivity (EffCS) — the equilibrium increase in global mean surface temperature from a doubling in atmospheric $CO_2$ — with several CMIP6 models having EffCS above the assessed likely and very likely ranges in the IPCC's 6$^{th}$ assessment report (AR6)[8]. A recent study (Hausfather et al., ref. [9]) suggested such "hot models" warm faster than is realistic, leading to concern that impact studies which draw from the full CMIP6 ensemble could overestimate the impacts of climate change. The IPCC tackled this by constraining projections of global mean temperature based on multiple lines of evidence[2]. This approach is not yet available for other quantities. Hausfather et al.[9] recommend filtering out models with EffCS outside the assessed likely range when performing impact assessment. While this may be reasonable for global mean surface air temperature, it is less clear what it means for assessments of regional change, and for quantities other than surface temperature. Projections of regional climate change, especially related to hydrological processes, often have a greater range than projections of global mean temperature. While this range is partly related to spread in the magnitude of the warming response to $CO_2$, changes in circulation patterns and heterogeneous forcings, such as aerosols[10] and land-use change[11], also contribute. Furthermore, some studies find high-EffCS models actually perform better than other models when evaluated for certain regions and applications[12,13,14].

We explore CMIP6-projected changes in regional climate impact drivers of fire, drought, and monsoon flooding, across a number of important regions, to assess the extent to which projected changes in these drivers are correlated with model EffCS. A regional climate impact driver is a measure of climate known to drive an important societal/environmental impact in the same region; the driver (e.g. extreme monsoon rainfall) can be directly diagnosed from an ESM, while the impact (e.g. flooding) typically cannot. Calculating changes at the end of the century relative to the recent past, following the Shared Socio-economic Pathway (SSP) 3-7.0 scenario[15] (see Methods), we find only very few examples of a significant correlation between projected changes in an impact driver and model EffCS. Even for the limited number of cases where a correlation with EffCS is identified, the spread of future change in a given impact driver across a single model ensemble (i.e. solely due to different model initial conditions), or between models with similar EffCS values, is comparable in magnitude to the spread across the full ensemble. This means that internal variability, as well as processes unrelated to EffCS, play at least as large a role as EffCS in determining future changes in our three climate impact drivers. Given the absence of any clear correlation between projected changes in our regional impact drivers and EffCS, we argue it is not justified to filter out ESMs for impact studies solely based on EffCS.

## Results

To demonstrate the effect of filtering out "hot models", we separate the CMIP6 ensemble into two sets: one with EffCS above 4.5K (referred to as high-EffCS hereafter), and another with EffCS below 4.5K (low-med-EffCS hereafter) and plot frequency distributions of the simulated change in each climate impact driver (referred to as metric hereafter), for the two model subsets, for each region of study. Applying such a distinction does not separate the frequency distributions of projected change in our three metrics, with the two distributions largely overlapping for the majority of metrics and regions (Fig. 1). In a few cases, the high-EffCS set includes values not projected by any of the low-med-EffCS models, for instance, large increases in high fire weather days in the Amazon region (AMZ), or in total rainfall in Central West Africa (CWAF). Even for these metrics and regions, the overlap between the two distributions is large, suggesting the projected change is not solely decided by EffCS. Moreover, high-EffCS models do not always project larger changes in our metrics; in Western North America (WNA), the high-EffCS set includes more simulations projecting a smaller increase in fire weather than the low-med-EffCS set. Similar conclusions emerge when just looking at the five CIMP6 models selected as core models (for the Inter-Sectoral Impact Model Intercomparison project, ISIMIP[16]) (see stars in Fig. 1).

Should the hottest models be filtered out to remove the most extreme projected changes? This would need a solid justification because filtering out particular models can result in vast ranges of potential impacts being excluded. For instance, in the five-member ISIMIP ensemble, removing the hottest model would ignore the upper half of the low-med CMIP6 spread of the flood metric in the core Indian monsoon (CIM) region, as well as the fire metric in AMZ (cf. red stars in Fig. 1). A necessary, though not sufficient, condition for constraining a model ensemble based on EffCS would be a statistically significant, ideally explainable, correlation between the metric (the projected change in the regional climate impact driver) and EffCS. In the following, we investigate whether such a correlation exists for our three metrics. We further assess how the spread in our three metrics, expressed as a function of EffCS, compares with the spread of the same metric for individual model ensembles, independent of EffCS, and across the full CMIP6 ensemble.

Table 1 shows, for each metric, the difference between the median value for the high-EffCS set and the median value for the low-med-EffCS set. It further shows the median and largest spread in each metric across only low-med-EffCS models. In all cases, except AMZ fire, the largest spread in the metric across low-med-EffCS models is bigger than the difference in the median of the metric between high- and low-med-EffCS sets. This is largely still the case when only the median spread in low-med-EffCS models is considered, although for CIM and CWAF, differences between the high- and low-med-EffCS ensembles are comparable to the spread in the low-med-EffCS set. This suggests that internal variability across an individual model or group of models clustered by EffCS, has a larger (or equivalent) influence on the metric than does EffCS. EffCS therefore appears to not be the sole cause of changes in our three selected metrics and thus also in the resultant impacts. We investigate these potential additional controls in the following subsections.

# Flood

We consider projected changes in the 20 most intense, 5-day precipitation accumulations; a metric indicative of the change in short-term flood risk (Methods). Most ensemble members show an increase in this metric over both CIM and CWAF regions, and either little change or a slight reduction over North (NCA) and South Central America (SCA) (Fig. 2). This agrees with previous studies[17], with an increase in the South Asian and West African monsoons attributed to the northern hemisphere land mass warming more rapidly than adjacent oceans, and more rapidly than the southern hemisphere due to the different land mass distributions[18]. In addition, atmospheric water vapour increases strongly with warming temperatures[19], increasing the availability of water in northern hemisphere monsoon circulations.

To test the relationship between our monsoon metric and EffCS across the full CMIP6 ensemble, we calculate a linear fit between the two quantities for every possible combination of one member from each model (Methods). For CIM, the relationship is positive and statistically significant at the 95% level for the majority (~83%) of combinations, while CWAF shows only ~6% of combinations with a significant correlation (Table 2). Both NCA and SCA show no significant correlation. In other words, the monsoon metric is correlated with EffCS in CIM, weakly correlated in CWAF, and not correlated at all in the other two regions. Even for CIM, some high-EffCS members project a smaller metric change than some of the low-med-EffCS members (Fig. 2). Moreover, the spread across members of an individual model (with the same EffCS value) is comparable to the median difference between low-med- and high-EffCS models, even for the CIM region (Fig. 2 and Table 1).

A study[20] evaluating the quality of simulated present-day monsoons (1979-2014) in 24 CMIP6 models found the ten best performing models to have EffCS values between 2.5K and 5.5K; the five best models are UKESM1-0-LL (EffCS 5.36K), CESM2 (5.16K), CESM2-WACCM (4.68K), MIROC6 (2.6K) and NorESM2-MM (2.49K). The spread in EffCS across the five models suggests little relationship between the quality of present-day simulated monsoon and model EffCS. The similarity of performance for CESM2 and NorESM2-MM is particularly interesting as these models share largely the same atmosphere and land models, but have very different ocean models. Their EffCS values, derived from the CMIP6 Abrupt-4x$CO_2$ experiment, are radically different because of differences in ocean heat uptake in the Southern Ocean[21]. A comparably accurate simulation of the present-day monsoon in these two models suggests their common atmosphere and land models are likely most important, while processes responsible for the radically different EffCS in the two models are of secondary importance with respect to representing the present-day monsoon.

Our results suggest the relationship between the monsoon metric and EffCS is weakened by other drivers. Aerosols are the primary non-EffCS controls on monsoon precipitation[22]. Increased aerosol loading increases atmospheric solar reflectivity close to the emission source[23], cooling the land surface more than the adjacent ocean and thereby weakening the land-ocean thermal gradient that drives the monsoon. Cooling, from the increased solar reflection, also decreases atmospheric water vapour, while increased absorption of solar radiation in the lower atmosphere, predominantly over land, increases atmospheric stability[23]. Both changes act to decrease monsoon precipitation. Modelling studies[24,25] suggest past observed (decreasing) trends in mean South Asian monsoon precipitation arise from the impact of anthropogenic aerosol emissions (decreasing precipitation) outweighing the impact of $CO_2$-forced warming (increasing precipitation). One study[26] shows that variations in regional aerosols are up to 4 times more efficient in impacting South Asian monsoon rainfall than equivalent variations in atmospheric $CO_2$, with extreme precipitation intensities particularly impacted by aerosols.

Aerosols also play an important role in the West African monsoon[27,28]. During the 1950s to 1970s, increasing North American (NA) aerosols induced a drying effect on Sahel rainfall. This effect was mediated by slower sea surface temperature (SST) responses to regional aerosol emission trends across the globe, with cooling of the tropical west Pacific driving a remote wetting signal over the Sahel and cooling of the tropical Atlantic inducing a drying signal. The combination of these three drivers saw Sahel rainfall decline over the 1950s to 1970s. During the 1970s to 2000s increasing African aerosol emissions drove a direct drying over the Sahel. This local effect was balanced by a wetting signal associated with Atlantic warming (due to decreasing NA aerosols), amplified by a continued remote wetting signal from the west Pacific. The net result was an increase in precipitation over this period.

For both the South Asian and West African monsoon, trends in regional aerosols have significant and competing impacts on precipitation. These impacts are not considered in the calculation of EffCS, which results solely from an increase in atmospheric $CO_2$ within a constant pre-industrial aerosol state. Any relationship between model EffCS and projected changes in monsoon precipitation, derived from more realistic future emission scenarios that include time-varying $CO_2$, aerosol, and aerosol precursor emissions, will be significantly weakened because of the confounding influence of aerosols.

Monsoons are also influenced by remote atmospheric teleconnections. The South Asian monsoon is influenced by the El Niño-Southern Oscillation (ENSO) SST variability[29]. The West African monsoon is influenced by multi-decadal Atlantic SST variability[30] and by atmospheric subsidence induced over the Sahara as a remote response to the South Asian monsoon[31]. In addition, the NCA and SCA monsoons are heavily impacted by ENSO variability and by anomalous subsidence forced by the West African and South Asian monsoons[32]. These impacts are sufficiently large that neither the NCA nor SCA monsoon metric show any relationship with EffCS across the CMIP6 ensemble.

In addition to regional aerosols and atmospheric teleconnections, monsoon precipitation also depends on local convection and atmospheric water availability, which themselves are sensitive to model representation of local land-vegetation-atmosphere interactions[33], the Himalayan snowpack (for CIM), and meso-convective-scale processes generally not resolved in ESMs[34]. The extent to which ESMs capture these mechanisms has little relation with EffCS. For reliable estimates of future monsoon rainfall, in addition to simulating the impact of increasing $CO_2$ (and other greenhouse gases), models also need to accurately simulate the impact of regional aerosol emissions, remote drivers of monsoon variability and small-scale, local processes and process interactions. All of these demands significantly weaken any link between model EffCS and projected changes in monsoon precipitation when realistic emission and land-use scenarios are used.

## Drought

Fig. 3 shows the projected change in the number of drought events per year as a function of EffCS (see methods for more detail). In almost all projections, the number of droughts increases in the future, consistent with earlier studies using CMIP5[35] and CMIP6 data[36,37]. However, for all regions and cases we fail to find a significant correlation between the drought change metric and EffCS (Table 2). Furthermore, the difference in the median metric between high- and low-med-EffCS models is always considerably smaller than the mean of the spread across only low-med-EffCS members (Table 1).

This is likely because changes in drought are forced by several climate drivers, with only one of these drivers being the magnitude of global warming. Local land warming, linked but not directly proportional to global mean warming, leads to increased evapotranspiration and vegetation water use, increasing drought risk, while higher atmospheric $CO_2$ counters this through increased plant water-use efficiency[38], making drought sensitive to local land-vegetation-atmosphere feedbacks. Many drought-prone regions today are located at the poleward edge of the subtropics, under the descending branch of the Hadley Cell, which induces persistent anticyclonic descending air, with dry conditions and high solar radiation. Future changes in regional drought will be highly sensitive to any systematic changes in the Hadley Cell. While observations indicate the Hadley cell has expanded poleward over recent decades[39] and models suggest a continued expansion in the future[40], zonal asymmetries in ocean warming, as well as differential responses between the ocean and land, lead to regional variations in projected Hadley cell expansion[41] and its impact on drought.

In addition, drought prone regions in the mid-latitudes are influenced by transient weather systems that propagate along the jet stream. Periodically, these systems are "blocked" by long-lived meanders in the jet[42]. Such blocking events are important drivers of drought[43,44]. The frequency and duration of such blocking depends on the strength of the jet, with a weaker jet allowing more and longer meanders (blocking events[45]). The jet strength is controlled by the tropical to pole tropospheric temperature gradient, with a stronger gradient resulting in a stronger and more zonal jet and fewer blocking events. The zonal mean response to increasing GHGs is warming of the tropical mid to upper troposphere and warming constrained close to the surface in polar regions, particularly in the Northern Hemisphere. As a result, it is thought that the jet stream will both strengthen and move poleward in the future[46]. Such a response would decrease the number and intensity of blocking meanders and change their preferred location: poleward and, over the Eurasian sector, eastward[47], with important consequences for future regional drought. While models have improved in simulating blocking, significant biases remain, even in the latest CMIP6 models[48]. Nevertheless, models show a consistent reduction in future blocking events, with a poleward and eastward shift of the main centres of action, potentially countering the impact of Hadley Cell poleward expansion. In addition to these dynamical/circulation controls on regional drought, trends in regional aerosol emissions can also impact atmospheric circulation and drought[49].

The competing impacts of increasing $CO_2$, trends in aerosol emissions, changes in regional circulation features, such as atmospheric blocking and regional land-atmosphere interactions, all mean accurately simulating past and future drought is a challenge. CMIP6 models reproduce historical drought with reasonable accuracy[37,50], although no single model stands out as best[51]. Additionally, while thermodynamic factors are responsible for average changes in the hydrological cycle, variability across models is governed by the dynamical (circulation) response to warming[52]. Based on our results, filtering models solely on EffCS is not justified for drought assessment.

# Fire

Fig. 4 shows that for almost all ensemble members, fire weather days are projected to increase in the future compared to today as a function of EffCS. However, in Fig. 1, fire metric frequency distributions for high- and low-med-EffCS models show a high degree of overlap, particularly for AUS. Over AMZ there is a clear tendency for high-EffCS models to simulate a larger increase in the number of fire weather days, while over WNA the opposite is the case.

Considering the full CMIP6 ensemble, a statistically significant relationship between EffCS and our fire metric can be seen for AMZ (Fig. 4a and Table 2). However, this relationship is heavily influenced by the very lowest and highest EffCS models that simulate the smallest and largest change in fire metric respectively, while models between these two extremes have a more nuanced and flatter relationship. For AUS, the relationship with EffCS is only partly significant (Table 2), and the largest increase in fire weather is simulated by the ACCESS model, which has an intermediate EffCS value (Fig. 4b). We also find that the spread within and across the high-EffCS models is within the range of the low-med-EffCS set (Fig. 1 and Fig. 4b), and therefore excluding them brings no clear benefit while reducing the sample size, or biasing the subsample, for impact studies. Similarly, the difference in behaviour between high- and low-med-EffCS median values is smaller than the average spread across low-med-EffCS models (Table 1). For WNA, there is no significant correlation with EffCS (Fig 4c and Table 2). Here, models with an EffCS around 3K project the largest increase, and the model with the highest EffCS (CanESM) projects the smallest change.

Wildfires pose a significant threat to communities and ecosystems. Climate influences fire through several mechanisms, such as fuel load, dryness of fuel from water deficit, heat, and spreading by winds. These factors make fire potentially sensitive to climate change, where higher temperatures or longer dry spells[53] can contribute to increased fire activity. Increased warming leads to higher evapotranspiration, which dries out fuels including vegetation, soils and litter, and fires are often linked with periods of drought. A strong relationship might be expected between projected changes in fire and EffCS, given that temperature is a key driver of fire weather. Yet our results for three fire-prone regions suggest EffCS is not the leading determinant of future change in fire. The results show a stronger relationship between EffCS and our fire metric in AMZ, where historic changes in fire regimes have been attributed to anthropogenic activity[54,55]. Yet we find no relationship in WNA, despite already observed increases in present-day fire weather which have been attributed to climate change[56,57]. In Australia, there has been some observed increase in fire weather[58] and fires may be becoming larger and more intense[59], although whether this is anthropogenically forced is less clear due to natural variability from the Indian Ocean Dipole and ENSO[60,61]). These results indicate that changes in regional fire weather are generally not directly related to EffCS. In reality, multiple interacting factors influence fire occurrence, not limited to temperature, but also fuel availability, dryness, natural fire ignition (e.g. by lightning), and ignition and suppression by humans[62,63]. Large scale modes of variability such as ENSO also influence fire weather from year-to-year, and warming of the Tropical North Atlantic Ocean has been associated with increased drought and fires in Amazonia[64]. As with floods and drought, an approach that is more nuanced than only focussing on EffCS is required before models can be excluded.

# Discussion

For ten regions and three climate impact drivers of extreme monsoon rainfall, drought, and fire we find no universal correlation between projected changes in these drivers and EffCS. For some regions and drivers, a correlation does exist (e.g. AMZ fire or CIM monsoon rainfall), while for most there is no clear relationship. Where a correlation is identified, it is not straightforward to attribute future changes in our regional metric solely to EffCS, with many other factors also contributing. For all three drivers, even the sign of the correlation depends on the region. Furthermore, for all drought regions, the majority of monsoon flood, and fire regions, the spread in projected changes across the full CMIP6 ensemble, or often across a single model ensemble, is larger than, or of similar magnitude to, the difference in the median value of the metric between high- and low-med-EffCS models. Considering the variety of regions and drivers, this should not be a surprise. All three drivers are influenced by multiple factors. For a few regions and drivers EffCS may be the most important (though not the only) controlling factor, but for most drivers and regions it is just one of many controls. While our results are contingent on the choice of region and driver studied, all regions are known hotspots for their respective impact. WNA, AMZ and AUS have experienced extreme recent wildfire years; CNA, MED and EAS strong droughts; and the South Asian monsoon major flood events in 2022 and 2023[65]. Moreover, here we study the climatic drivers of regional impacts. Impact models themselves (as well as real-world societal impacts) are more complex, involving numerous interactions between climatic drivers, biogeophysical and human responses and subsequent societal impact. We therefore believe the results from impact models will be even less clearly related to EffCS than the climate impact drivers studied here.

Societal impacts of climate change often occur at small spatial scales. As scales decrease, uncertainty in the forced climate change signal increases, partly because natural climate variability also increases. It is therefore even more important that information sampling this uncertainty is not thrown away without good reason. Thresholds may exist that trigger societal impacts, considering a broad range of possible regional climate changes is therefore important for understanding the true risk to society. Rejecting ESM projections for regional impact assessment, based solely on EffCS, should therefore be avoided in the absence of additional motivation, especially because it risks ignoring large parts of the range of potential impacts. It is important to check if there is a statistically significant relationship between the criterion used for model selection (e.g. EffCS) and the climate drivers of the impacts in question. If this is not the case, or only partially true, then other factors should be considered before rejecting, or weighting models[66,67]. For three impact drivers, over several regions, other factors such as aerosol forcing, atmospheric circulation and teleconnection changes, as well as local atmosphere-land-vegetation interactions are often as important as EffCS for determining the projected change in a regional impact driver. For each driver, it is important to identify the main processes inducing significant change in the driver and carefully evaluate each model's representation of these processes. . It is not sufficient to assume such regional responses simply scale with model EffCS, as they generally do not.

Our results emphasise the important role of internal (or natural) variability plays in changes in regional climate drivers and their resultant impacts. For most drivers we studied, the spread across a single model ensemble is often larger than the difference in the driver between high-EffCS and low-med-EffCS models. This is also true when a smaller subset of models is considered, for example those selected for the ISIMIP3 activity (Supplementary Fig. S1). Sampling a sufficient range of plausible impacts, including possible extreme outcomes, requires impact model experiments sample the full range of plausible changes in regional climate drivers. When it is not practical to run an entire CMIP ensemble (including all model members) through an impact model, analysis of relevant climate impact drivers across the full multi-model ensemble can help inform the selection of ESMs, and model members, that both spans the range of plausible future change, and emphasizes the most likely outcomes.

Selecting ESMs for regional impact studies based solely on EffCS has little (often zero) justification. Without more careful consideration, any such reduced ensemble used for impact studies will be unnecessarily biased. In the worst case, this will lead to the exclusion of models with plausible estimates of regional impact drivers, and the impacts themselves, with negative consequences for the robustness of scientific support for decision-making.

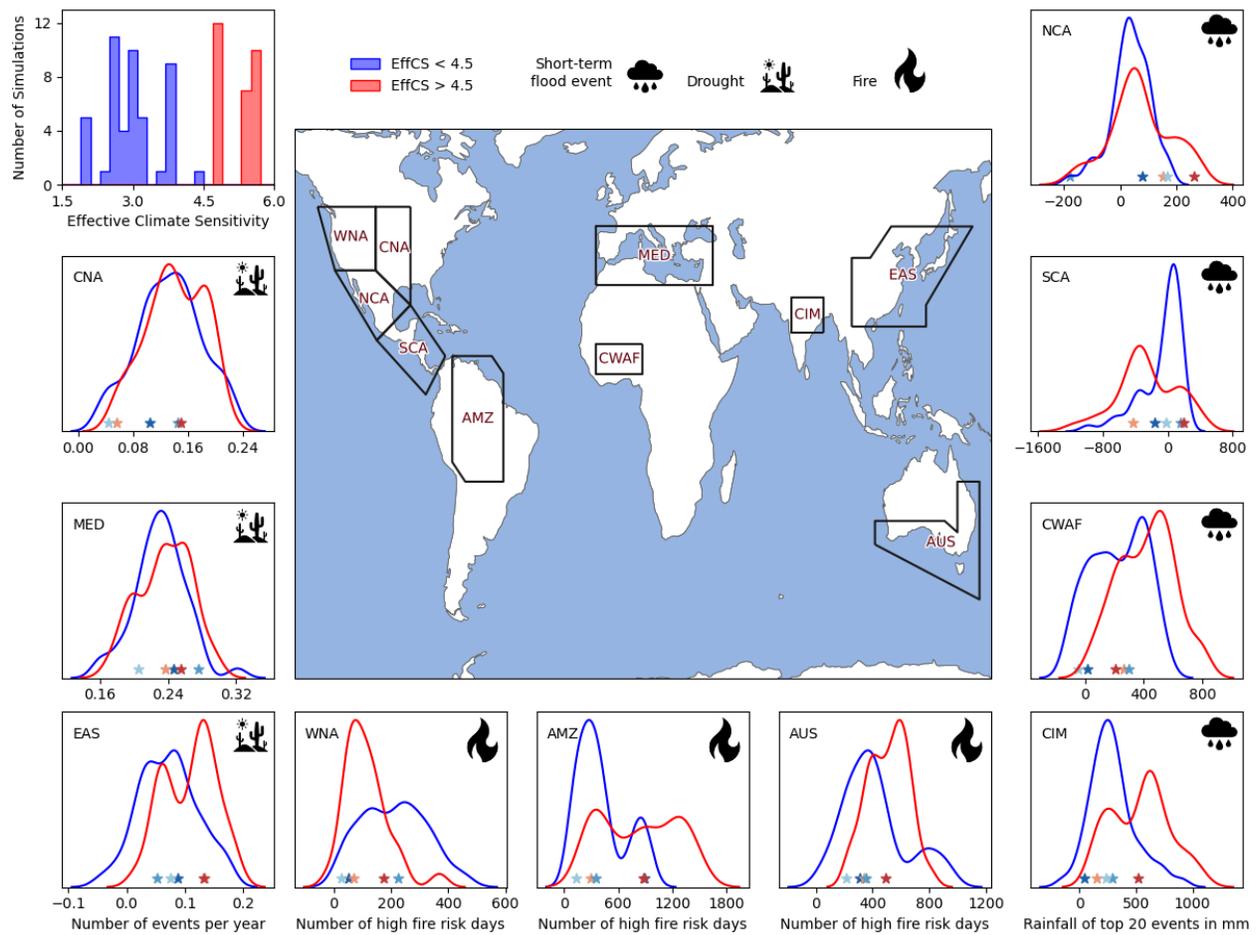

**Figure 1: Frequency distribution of projected change in our three regional climate drivers (metrics).** Top left: number of model simulations with a given EffCS value below (blue) or above (red) 4.5K. Other panels show the distribution of projected change in the metrics for - drought, left; fire, bottom; monsoon flood, right - for each of the two model subsets (vertical axis shows normalised occurrence, i.e. the area under each curve is equal to 1). The value of 4.5K is between the upper bounds of the AR6 assessed likely (4K) and very likely (5K) ranges for EffCS, and provides for a sufficient number of simulations in each of the subsets. Stars indicate the metric values for five model simulations that are used as input to many impact models[16], and are colour-coded by EffCS (cf. legend of Fig. 2). The map shows the regions over which the impact drivers are analysed. See Methods section for more detail on how the projected change in each driver was calculated, model simulations, and statistical methods.

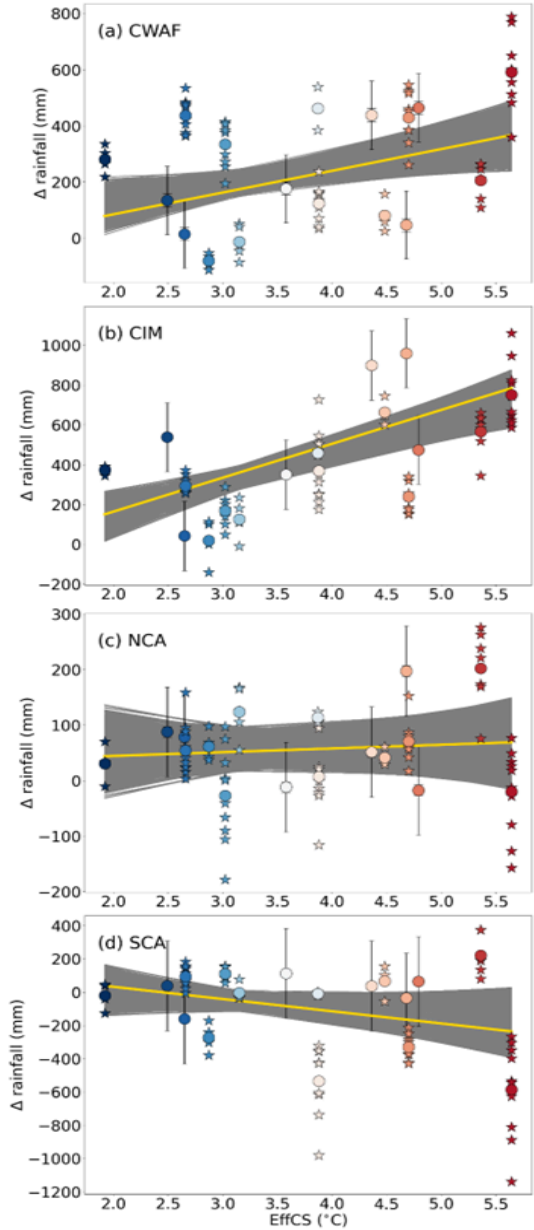

**Figure 2: Change in area-averaged, total cumulative 5-day rainfall of the 20 most intense rainfall events per 20-year period, versus model EffCS value.** Change is calculated as 2081-2100 (under SSP3-7.0) minus 1995-2014. (a) Core West African monsoon region (CWAF),

(b) Core Indian monsoon region (CIM), (c) North Central America (NCA), (d) South Central America (SCA). For each model, individual ensemble members are denoted by stars, and the ensemble mean value by a circle. Models with a single ensemble member include error bars indicating the largest (in black) and smallest (in grey, may be hidden behind model symbol) standard deviations as derived from all models with multiple ensemble members. The yellow line shows the best fit for the first (in some cases, only) ensemble member for each model. The grey area is formed from numerous individual lines each showing the best fit to a possible combination of one ensemble member from each model. The model EffCS value is mapped to colour and the five model simulations that are used as input to many impact models[16] are highlighted in bold (see legend).

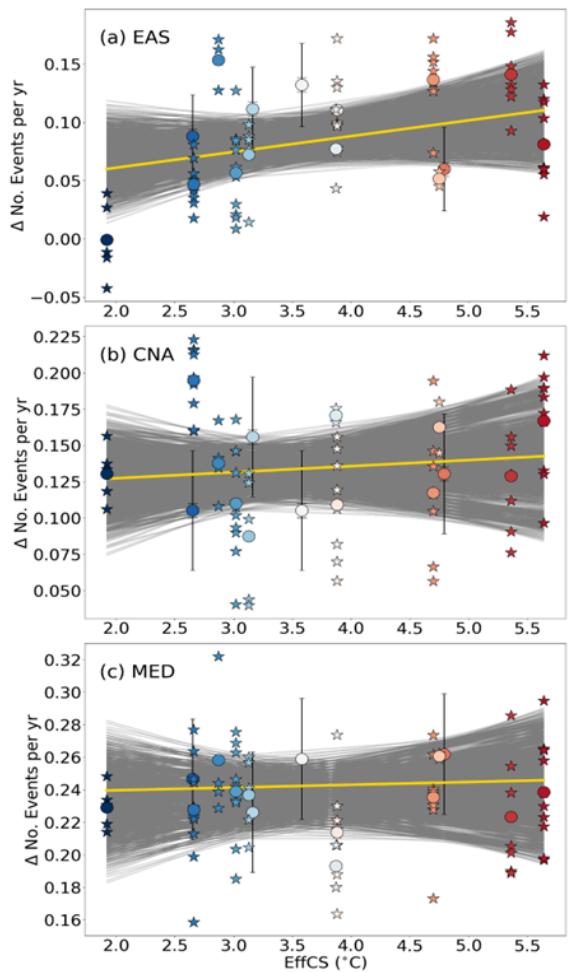

**Figure 3: As Fig. 2 but for change in the number of extreme drought events per year.** (a) East Asia (EAS), (b) Central North America (CNA), (c) Mediterranean (MED).

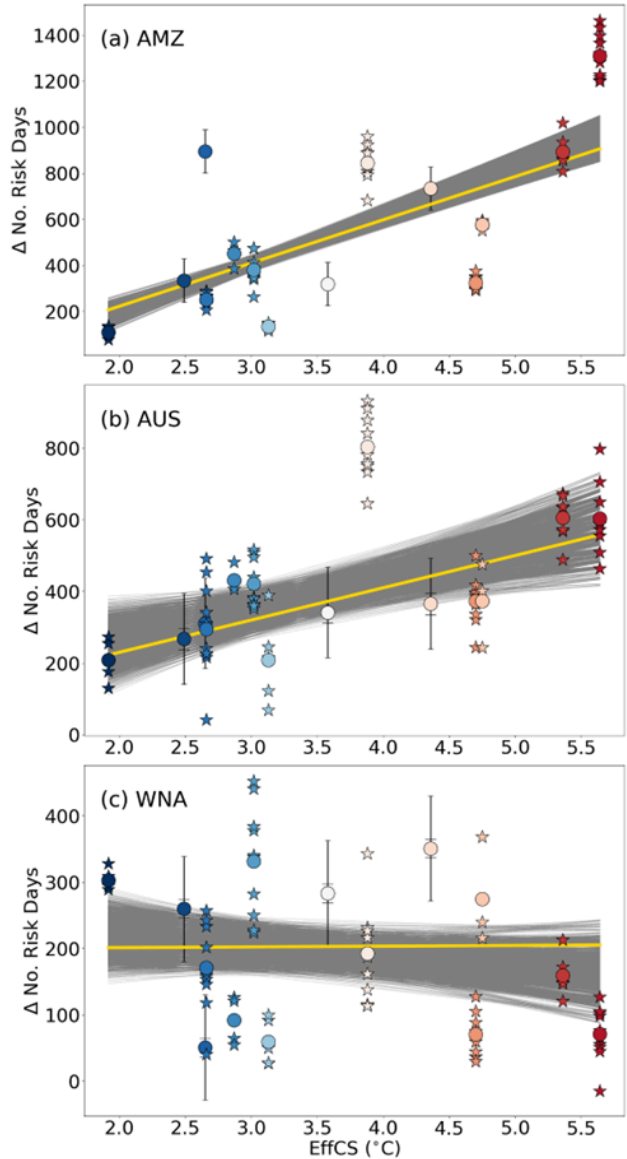

**Figure 4: As Fig. 2 but for change in the number of fire risk days per 20 year period.** (a) Amazon (AMZ), (b), Australia (AUS), (c) Western North America (WNA).

Table 1. Difference in median impact metric values between high-EffCS and low-med-EffCS models, compared with the spread across members for individual low-med-EffCS models.

| Region | Difference between high- and low-med-EffCS medians (absolute value) | Average spread in low-med-EffCS models (absolute value) | Largest spread in low-med-EffCS models (absolute value) |
|---|---|---|---|
| **Flood** | | | |

| Core West African Monsoon (CWAF) | 216.80 mm | 149.12 mm | 219.68 mm |
|---|---|---|---|
| Core Indian Monsoon (CIM) | 329.46 mm | 201.96 mm | 550.94 mm |
| North Central America (NCA) | 30.77 mm | 119.86 mm | 275.63 mm |
| South Central America (SCA) | 339.18 mm | 209.01 mm | 657.46 mm |
| **Drought** | | | |
| East Asia (EAS) | 0.044692 | 0.079403 | 0.118613 |
| Central North America (CNA) | 0.0056226 | 0.071512 | 0.127481 |
| Mediterranean (MED) | 0.0074561 | 0.075394 | 0.118434 |
| **Fire** | | | |
| Amazon (AMZ) | 523 days | 130 days | 278 days |
| Australia (AUS) | 129 days | 240 days | 450 days |
| Western North America (WNA) | -122 days | 143 days | 230 days |

Table 2: Significance of correlation (at 95% significance level) between change in impact drivers and EffCS.

| Region | Number of low-med-EffCS members exceeding high-EffCS median (models contributing those members) | Percentage of statistically significant correlations in low-med-EffCS models | Percentage of significant correlations in all models | r1 ensemble member correlation | Highest ensemble member of low-med-EffCS models correlation |
|---|---|---|---|---|---|
| **Flood** | | | | | |
| Core West African Monsoon (CWAF) | 6(2) | 0 | 6.25 | Not significant | Not significant |
| Core Indian Monsoon (CIM) | 5 (3) | 42.97 | 83.54 | **Significant** | **Significant** |
| North Central America (NCA) | 18 (9) | 0 | 0.34 | Not significant | Not significant |
| South Central America (SCA) | 41 (11) | 0 | 0 | Not significant | Not significant |
| **Drought** | | | | | |
| East Asia (EAS) | 9 (4) | 21.09 | 3.857 | Not significant | Not significant |
| Central North America (CNA) | 21 (7) | 0 | 4.248 | Not significant | Not significant |
| Mediterranean (MED) | 20 (8) | 0 | 4.346 | Not significant | Not significant |
| **Fire** | | | | | |
| Amazon (AMZ) | 5 (2) | 0 | 100 | **Significant** | **Significant** |
| Australia (AUS) | 11 (2) | 1.56 | 43.55 | **Significant** | Not significant |
| Western North America (WNA) | 38 (8) | 0 | 0 | Not significant | Not significant |

# Methods

## Climate impact drivers

We use drivers representing three different climate impacts: floods, drought and fire. All three impacts are clearly associated with climate change and expected to worsen under future conditions[68,69]. For each climate impact driver, we identify a measure whose magnitude can be expected to be representative of the magnitude of the associated impact, and analyse the change in this measure between a historical period (1995-2014) and a future period (2081-2100) under the SSP3-7.0 scenario. SSP3-7.0 is a high warming scenario from Tier 1 of the Scenario Model Intercomparison Project (ScenarioMIP[70]) that can be expected to inform many climate impact studies in the near future; it is is part of the current simulation round of the Inter-Sectoral Impact Model Intercomparison Project (ISIMIP, www.isimip.org).

### Flood

As a proxy for pluvial flood risk associated with heavy monsoon rainfall, we calculate the area averaged cumulative five-day rainfall amounts for a given region for the historical and future periods. The difference between the total rainfall amounts of the top twenty such rain events from each period is then calculated as our flood metric.

### Drought

Drought is defined as an *anomalous condition* with respect to local and seasonal characteristics, rather than an absolute threshold. We use the Standardised Precipitation Evapotranspiration Index (SPEI)[71,72], with the Hargreaves approximation for potential evapotranspiration[73] (PE) to investigate future change in extreme drought occurrence. SPEI is a widely used metric to characterise atmospheric drought and integrates not only precipitation but also evapotranspiration, meaning that it accounts for the role of temperature in determining the amount of water the atmosphere holds, which is important for our study given that we look at the role of climate sensitivity expressed by temperature.

We follow ref. [74] to calculate the number of drought events per year based on SPEI. For each ensemble member we calculate the difference in average number of drought events for each region between the future and historical periods. We use v1.8 of the SPEI.R library[75] to calculate SPEI. Selecting an accumulation period of six months, we use a threshold of months with SPEI < -2 to identify extreme droughts. SPEI is calculated by fitting a log-logistic distribution to the difference between precipitation and PE for monthly data. The average SPEI value is 0, which corresponds to a cumulative probability of 50% for this difference, whilst the standard deviation is 1. Following a normal distribution, a threshold of < -2 (two standard deviations lower than the average) will find extreme droughts in 2.3% of all months.

## Fire

Wildfires are key drivers of ecosystem dynamics and carbon cycling[76] and, while overall global burned area is decreasing[77], climate change driven increases in frequency and intensity of wildfires have strong effects on human health[78,79] and economic value of land[80]. We calculate the number of fire-risk days for each period, using the Canadian Fire Weather Index (FWI). The FWI integrates effects of local weather conditions and fuel moisture on fire behaviour. It can be calculated from climatic variables alone (temperature, precipitation, relative humidity and wind speed) and, because the different sub-indices can be regionally calibrated, has been widely used across different forest types in the world. Established by[81] the FWI has seen some development[82] and is widely used to gauge fire risk. FWI is calculated via a multistep process taking many factors of forest fire risk into account and culminating in a daily index with arbitrary units. These numbers are then typically classified according to chosen thresholds into several classes of risk, such as low, medium, and high; resulting in a risk class for every day at a given point[82]. For our purposes of evaluating the impact of climate change as predicted by different climate models, we employ a simplified scheme with a single threshold. We classify all days at a given point with a FWI above that threshold as high fire risk days and count the number of such days in a 20 year period for a given region. By comparing this number for the future period with the historical period, we obtain a measure for the change in fire risk induced by climate-driven changes in weather.

# Choice of regions

The three climate impact drivers are each evaluated over a set of three (four for flood) regions where the corresponding impact is known to be large and to affect large numbers of people. Regions are further selected to cover different continents and both the tropics and temperate mid-latitudes.

For flood, we looked at four different regions: Core Indian Monsoon (CIM), Central West African Monsoon (CWAF), North and South Central American Monsoons (NCA and SCA). The CIM region was defined as the area between 18 – 27 N and 74 – 88 E; the CWAF to be 7.5 – 15 N and 10 W – 10 E as defined in[83] and the NCA and SCA regions as defined in the AR6 WGI Reference Set of Land and Ocean Regions[84,85].

The regions selected for drought are also defined in the AR6 WGI, which are the Mediterranean (MED), considered a climate change hotspot, and East Asia (EAS) and Central North America (CNA), which are agriculturally important regions affected by severe droughts in recent years.

For fire, we choose three different regions that are known to suffer from forest fires and that are at high risk from a future increase in this impact, namely Western North America (WNA) as given by the IPCC AR6 regions, the Amazon basin (AMZ), here defined as the combination of two AR6 regions (South-American-Monsoon and Northern South-America), as well as a region of Australia (AUS), again defined by the combination of two AR6 regions (Eastern Australia and Southern Australia).

## Statistical Methods

The measure used to define Earth System Model (ESM) sensitivity in our analysis is the Effective Climate Sensitivity (EffCS) and not the equilibrium climate sensitivity. The true equilibrium climate sensitivity is rarely assessed for ESMs as this requires very long model integrations. We use the same EffCS values as ref. [9], calculated as described in ref. [86] and consistent with what is reported in Chapter 7 of the Sixth IPCC Assessment Report[87].

We establish the relationship between our choice of metric for changes in different regional impacts such as short term flooding events, fire occurrence days or drought indices with the EffCS of ESMs by calculating a linear fit between the two quantities. In this, we assume that a linear relationship exists between the two quantities. We then determine the statistical significance of this relationship using null hypothesis testing based on the Student's t-test for significance in the 95% confidence interval. When considering metrics of ensemble projections from a single model, we use the minimum and maximum ensemble member metric values from models with ensemble sizes larger than one or just the single ensemble member available and calculate lines for all possible combinations of these metric values across models (see Figs. 2-4). We then calculate the number of statistically significant lines using null hypothesis testing as before and report results. Hence, we sample the full range of possible metric values.

## Data and Code Availability

Our study used a selection of eighteen different Earth System Models from the Sixth Coupled Model Intercomparison Project (CMIP6). Our selection criteria were based on maximising diversity across models and EffCS values of models. For each model that had the requisite variables for calculating a given metric, we used the first ten or all available ensemble members (if fewer than ten were available) to represent the spread due to internal variability. All data was downloaded from Earth System Grid Federation (ESGF) nodes (https://esgf.llnl.gov/) and preprocessed using the open source software ESMValTool (https://esmvaltool.org/)[88]]. The list of models and ensemble members used in our work is provided in Table 3. The r1 member is always included; this is also the ensemble member that is used in many impact studies relying on single-member multi-model ensembles[89]. Not all models had the necessary variables for every climate impact driver, therefore the set of models differs slightly between Figures 2-4. Software developed to calculate the three impact metrics, perform statistical analyses and make our plots has been archived, together with our preprocessed data, at https://doi.org/10.5281/zenodo.10533860.

Table 3: List of CMIP6 models and ensemble members used for the three climate impact drivers along with their EffCS values as given in ref. [9].

| Model | EffCS | Ensemble members for | Ensemble members for | Ensemble members for |
|---|---|---|---|---|

|  |  | Monsoon impacts | Drought impacts | Fire impacts |
| --- | --- | --- | --- | --- |
| ACCESS-ESM1-5 | 3.88 | r1,r2,r4-10 (9 variants) | r1,r2,r4-10 (9 variants) | r1,r2,r4-10 (9 variants) |
| AWI-CM-1-1MR | 3.16 |  | r1 |  |
| CESM2-WACCM | 4.68 | r1 |  |  |
| CMCC-ESM2 | 3.58 | r1 | r1 | r1 |
| CNRM-ESM2-1 | 4.79 | r1 | r1 |  |
| CanESM5 | 5.64 | r1-10 (10 variants) | r1-10 (10 variants) | r1-10 (10 variants) |
| EC-Earth3-AerChem | 3.87 | r1, r3 (2 variants) | r1, r3 (2 variants) |  |
| FGOALS-g3 | 2.87 | r1,r3-5 (4 variants) | r1,r3-5 (4 variants) | r1,r3-5 (4 variants) |
| GFDL-ESM4 | 2.65 | r1 | r1 | r1 |
| INM-CM5-0 | 1.92 | r1-3 (3 variants) | r1-5 (5 variants) | r1-5 (5 variants) |
| IPSL-CM6A-LR | 4.7 | r(1-9) (9 variants) | r(1-9) (9 variants) | r(1-9) (9 variants) |
| KACE-1-0-G | 4.75 | r1-3 (3 variants) | r1, r2 (2 variants) | r1-3 (3 variants) |
| MIROC-ES2L | 2.66 | r1-10 (10 variants) | r1-10 (10 variants) | r1-10 (10 variants) |
| MPI-ESM1-2-HR | 2.98 | r(1-10) (10 variants) | r(1-10) (10 variants) | r(1-10) (10 variants) |
| MRI-ESM2-0 | 3.13 | r1-5 (5 variants) | r1-5 (5 variants) | r1-5 (5 variants) |
| NorESM2-MM | 2.49 | r1 |  | r1 |
| TaiESM1 | 4.36 | r1 |  | r1 |
| UKESM1-0-LL | 5.36 | r(1,2,3,4,8,9,10) (7 variants) | r(1,2,3,4,8,9,10) (7 variants) | r(1,2,3,4,8,9,10) (7 variants) |

Inter-Sectoral Impact Model Intercomparison Project (ISIMIP2b). *Geosci. Model Dev.* **10**, 4321–4345 (2017).
## Acknowledgements

RS, CJ and AGT are funded by UKRI-NERC TerraFIRMA (NE/W004895/1). JW, RAB, CB and CDJ are supported by the Met Office Hadley Centre Climate Programme funded by DSIT. KZ, CJ and CDJ acknowledge funding from the European Union Horizon 2020 project ESM2025, grant number 101003536. MM acknowledges funding from the Horizon Europe project OptimESM, grant number 101081193. KW is supported through the project S1 ("Diagnosis and Metrics in Climate Models") of the Collaborative Research Centre TRR 181 "Energy Transfers in Atmosphere and Ocean" funded by the Deutsche Forschungsgemeinschaft (DFG, German Research Foundation) - Projektnummer 274762653. KW's research for this study was also funded by the Deutsche Forschungsgemeinschaft (DFG, German Research Foundation) through the Gottfried Wilhelm Leibniz Prize awarded to Veronika Eyring (Reference number EY 22/2-1). Discussions about this article were supported by the COST Action CA19139 PROCLIAS (PROcess-based models for CLimate Impact Attribution across Sectors), supported by COST (European Cooperation in Science and Technology; https://www.cost.eu). We acknowledge high-performance computing support from the Centre for Environmental Data Analysis's JASMIN supercomputer in the United Kingdom, the German Climate Computing Centre (DKRZ) under project no. bd1083, and the National Supercomputing Centre (NSC) in Sweden. We thank the CMIP6 modelling groups for making their simulation outputs available through the Earth System Grid Federation.

## Author Contributions

CJ, CDJ, RB, MM, CPOR, and JS conceived the study. RS implemented the flood impact metric and statistical analyses, JW implemented the drought metric using previous work by KW, and KZ implemented the fire metric. CB, KW, AGT and MM contributed to the data analysis. CJ, RS and JS led the writing of the final paper. All authors contributed to the discussions, development of metrics and analyses, writing, proof-reading and editing of the manuscript.

## Competing Interests

The authors declare no competing interests.

# Corresponding Authors

Correspondence to Ranjini Swaminathan or Jacob Schewe.